\documentclass[12pt]{article}
\usepackage{cite}
\usepackage{CJK}
\usepackage{amssymb}
\usepackage{amsmath}
\usepackage{enumerate}
\usepackage{graphicx}
\usepackage{amsfonts}
\usepackage{url}
\usepackage{bm}

\usepackage{pifont}
\usepackage{epsfig,subfigure,dsfont,amsthm,amsbsy,mathrsfs,amscd}

\def\be{\begin{equation}}
\def\ee{\end{equation}}
\def\ba{\begin{array}}
\def\ea{\end{array}}
\input amssym.def

\newcommand\btd{\raise 2pt \hbox{$\hat\bigtriangledown$}\hskip 1.5pt}
\newcommand\bt{\raise 2pt \hbox{$\bigtriangledown$}\hskip 1.5pt}
\begin{document}
 \title{\large\bf Sum uncertainty relations for arbitrary $N$ incompatible observables}
\author{Bin Chen$^{1}$ \& Shao-Ming Fei$^{1,2}$$^\ast$\\[10pt]
\footnotesize
\small $^{1}$School of Mathematical Sciences, Capital Normal University, Beijing 100048, China\\
\footnotesize
\small $^{2}$Max-Planck-Institute for Mathematics in the Sciences, Leipzig 04103, Germany}
\date{}

\maketitle

\centerline{$^\ast$ Correspondence to feishm@cnu.edu.cn}
\bigskip

\begin{abstract}

We formulate uncertainty relations for arbitrary $N$ observables.
Two uncertainty inequalities are presented in terms of the sum of variances and standard deviations, respectively.
The lower bounds of the corresponding sum uncertainty relations are explicitly derived.
These bounds are shown to be tighter than the ones such as derived from the uncertainty inequality for two observables [Phys. Rev. Lett. 113, 260401 (2014)].
Detailed examples are presented to compare among our results with some existing ones.
\end{abstract}
\bigskip

Uncertainty principle, as one of the most fascinating features of the quantum world, has attracted considerable attention
since the innovation of quantum mechanics. The corresponding uncertainty inequalities are of great importance
for both theoretical investigation and experimental implementation.
In fact, the Heisenberg uncertainty principle \cite{Heisenberg,y2,y3} typically said
that measuring some observables on a quantum system will inevitably disturb the system.
There are many ways to quantify the uncertainty of measurement outcomes, for instance, in terms of the noise and disturbance \cite{y6,BLW1},
according to successive measurements \cite{y7,y8,y16,y17}, as informational recourses \cite{y9}, in
entropic terms \cite{y10,y11}, and by means of majorization technique \cite{y12,y13,y14}.
The traditional approach that deals with quantum uncertainties raised in many different
experiments uses the same pre-measurement state.
For a pair of observables $A$ and $B$, the well-known Heisenberg-Robertson uncertainty relation \cite{Heisenberg,Robertson}
says that,
\begin{equation}\label{UR}
\Delta A\Delta B\geq\frac{1}{2}|\langle\psi|[A,B]|\psi\rangle|,
\end{equation}
where $\Delta(\Omega)=\sqrt{\langle \Omega^{2}\rangle-\langle \Omega\rangle^{2}}$ is the standard deviation of an observable $\Omega$, and $[A,B]=AB-BA$.
Heisenberg-Robertson uncertainty relation implies the impossibility to determine the precise values of two non-commuting observables simultaneously.
However, the lower bound in the uncertainty inequality (\ref{UR}) can be trivial, even if the state $|\psi\rangle$ is not
a common eigenstate of the two observables. In fact, the product of the standard deviation $\Delta A\Delta B$ is null if the measured state $|\psi\rangle$
is an eigenstate of one of the two observables. Thus, the formulation of uncertainty relation in terms of product form of standard deviation has a
drawback in characterizing the incompatibility of the observables. To deal with such problems, uncertainty relations based on sum of variances
have been taken into account. Such sum uncertainty relations have very useful applications in quantum information theory,
such as entanglement detection \cite{Hofmann,G¨¹hne} and error-disturbance relation \cite{Busch}.
In \cite{Maccone} L. Maccone and A. K. Pati recently provided two stronger uncertainty relations in terms of the sum of variances.
It is shown that the lower bounds of their uncertainty inequalities are nontrivial, whenever the two observables are incompatible with respect to
the measured states (the states are not common eigenstates of both two observables).

Physically, besides pairs of non-commutating observables like position and momentum, there are also
triple non-commutating observables like the three component vectors of spin, angular moment or the isospin of particles.
Hence it is also important to find the uncertainty relations for a set of finite number of observables.
In deed, one can obtain an uncertainty relation for multiple observables by summing over the uncertainty inequalities for all the pairs of these observables.
However, the resulting lower bounds of such obtained uncertainty relation for multiple observables are generally not tight.

In this article, we explore the uncertainty relations for arbitrary $N$ incompatible observables.
We present a sum of variance-based uncertainty relations and a standard deviation-based sum uncertainty relation for $N$ observables.
The lower bounds presented in these inequalities are tighter than the one from summing over all the inequalities for pairs
of observables \cite{Maccone} and than the one in \cite{Pati}.
Our uncertainty relations are also useful in capturing the incompatibility among the $N$ observables: the relations are nontrivial
as long as the measured state is not a common eigenstate of all the $N$ observables.

\medskip
\noindent{\bf Results}

\medskip
{\sf Variance-based sum uncertainty relations}~
We first consider uncertainty relations based on the sum of variances of every observables and
the sum of standard deviation of pairs of observables:

{\bf Theorem 1}~For arbitrary $N$ observables $A_{1}$, $A_{2}$, $\ldots$, $A_{N}$, we have the following variance-based sum uncertainty relation:
\begin{equation}\label{cb1}
\begin{split}
\sum_{i=1}^{N}(\Delta A_{i})^{2}\geq&\frac{1}{N-2}\left\{\sum_{1\leq i<j\leq N}[\Delta(A_{i}+A_{j})]^{2}\right.\\
&\left.-\frac{1}{(N-1)^{2}}\left[\sum_{1\leq i<j\leq N}\Delta(A_{i}+A_{j})\right]^{2}\right\}.
\end{split}
\end{equation}

See Methods for the proof of Theorem 1.

To show that our bound (\ref{cb1}) is not a trivial generalization from uncertainty inequality for two observables,
let us consider the recent result in \cite{Maccone}, where the authors obtained an uncertainty inequality for two observables by
using parallelogram law in Hilbert space:
\begin{equation}
(\Delta A)^{2}+(\Delta B)^{2}\geq\frac{1}{2}[\Delta(A+B)]^{2}.
\end{equation}
From this inequality we can get an inequality for arbitrary $N$ observables $A_{1},A_{2},\ldots,A_{N}$.
Noting that
\begin{equation}\nonumber
\sum_{i=1}^{N}(\Delta A_{i})^{2}=\frac{1}{(N-1)}\sum_{1\leq i<j\leq N}[(\Delta A_{i})^{2}+(\Delta A_{j})^{2}],
\end{equation}
we have
\begin{equation}\label{tb1}
\sum_{i=1}^{N}(\Delta A_{i})^{2}\geq\frac{1}{2(N-1)}\sum_{1\leq i<j\leq N}[\Delta(A_{i}+A_{j})]^{2}.
\end{equation}
The right hand side of (\ref{tb1}) is a lower bound of variance-based sum uncertainty relation for $N$ observables.

To show that our new bound (\ref{cb1}) is tighter than (\ref{tb1}), it is sufficient to prove the following inequality for ${N(N-1)}/{2}$ positive numbers:
\begin{equation}
\begin{split}
\frac{1}{N-2}\left[\sum_{i=1}^{N(N-1)/2}x_{i}^{2}-\frac{1}{(N-1)^{2}}\left(\sum_{i=1}^{N(N-1)/2}x_{i}\right)^{2}\right]\\
\geq\frac{1}{2(N-1)}\sum_{i=1}^{N(N-1)/2}x_{i}^{2}.
\end{split}
\end{equation}
This inequality is equivalent to
\begin{equation}\label{a}
n\sum_{i=1}^{n}x_{i}^{2}\geq\left(\sum_{i=1}^{n}x_{i}\right)^{2}, ~~~n=\frac{N(N-1)}{2}.
\end{equation}
By taking into account that
$$
(n-1)\sum_{i=1}^{n}x_{i}^{2}=\sum_{1\leq i<j\leq n}(x_{i}^{2}+x_{j}^{2})\geq2\sum_{1\leq i<j\leq n}x_{i}x_{j},
$$
we have that the bound in (\ref{cb1}) is tighter than the one in (\ref{tb1}).

It is obvious that if the lower bound (\ref{cb1}) is zero, so is the bound (\ref{tb1}), and each $\Delta(A_{i}+A_{j})$ is equal to zero.
In this case the state $|\psi\rangle$ must be an eigenstate of each $A_{i}+A_{j}$, hence the common eigenstate of all $A_{i}$
(To see this, suppose that $|\psi\rangle$ is a common eigenstate of $A_{i}+A_{j}$, $A_{j}+A_{k}$ and $A_{i}+A_{k}$.
Then $|\psi\rangle$ is an eigenstate of $A_{i}+A_{j}+A_{k}$, thus the common eigenstate of $A_{i}$, $A_{j}$ and $A_{k}$).
That is to say, if the $N$ observables are incompatible associated with the state $|\psi\rangle$, then the lower bound (\ref{cb1}) must be nonzero.
For mixed state $\rho=\sum_{i}p_{i}|\psi_{i}\rangle\langle\psi_{i}|$, the lower bound (\ref{cb1}) is nontrivial as long as
there exits one (or more) $|\psi_{i}\rangle$ in the ensemble is not a common eigenstate of all $A_{i}$.
Therefore, the lower bound (\ref{cb1}) of sum variance-based uncertainty relation captures better the incompatibility
of arbitrary finite number of observables.

\begin{figure}
\centering
\includegraphics[width=7cm]{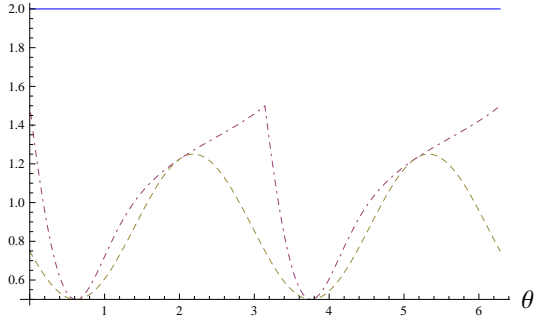}
\caption{The horizontal line is the sum of the variances $(\Delta X)^{2}+(\Delta Y)^{2}+(\Delta Z)^{2}$.
The dot-dashed line is the bound (\ref{cb1}), with the maximal value 1.5 attained at $\theta=0$ and $\pi$.
The dashed line is the bound (\ref{tb1}), with the maximal value 1.25.}\label{v1}
\end{figure}

As a detailed example, let us
consider the Pauli matrices $X=|0\rangle\langle1|+|1\rangle\langle0|$, $Y=-i|0\rangle\langle1|+i|1\rangle\langle0|$, $Z=|0\rangle\langle0|-|1\rangle\langle1|$
as the spin measurement operators on a qubit pure state with the density matrix given by
the Bloch vector $\overrightarrow{r}=(\frac{1}{\sqrt{2}}\cos\theta,\frac{1}{\sqrt{2}}\cos\theta,\sin\theta)$.
Then we have $(\Delta X)^{2}+(\Delta Y)^{2}+(\Delta Z)^{2}=2$, $[\Delta(X+Y)]^{2}=1-\cos2\theta$, and $[\Delta(Y+Z)]^{2}=[\Delta(X+Z)]^{2}=\frac{5}{4}+\frac{1}{4}\cos2\theta-\frac{\sqrt{2}}{2}\sin2\theta$.
The comparison between the lower bounds (\ref{tb1}) and (\ref{cb1}) is given in FIG \ref{v1}.
Apparently our bound is tighter than (\ref{tb1}).

\medskip
{\sf Standard deviation-based sum uncertainty relations}~
In this section, we formulate uncertainty relations in terms of sum of standard deviations.
For two observables $A$ and $B$, one can easily get an uncertainty inequality:
\begin{equation}\label{cb2}
\Delta A+\Delta B\geq\max\{\Delta(A+B),\Delta(A-B)\},
\end{equation}
since $\Delta(A\pm B)\leq\Delta A+\Delta B$ \cite{Pati}.
If the lower bound (\ref{cb2}) is trivial, then the measured state must be an eigenstate of both $A+B$ and $A-B$,
thus also a common eigenstate of $A$ and $B$. This implies that standard deviation-based sum uncertainty relations
are also useful in characterising the incompatibility of observables, namely, the lower bound (\ref{cb2}) is nonzero
if the two observables are incompatible associated to the measured state.
For arbitrary $N$ observables, we have the following conclusion:

{\bf Theorem 2}~For arbitrary $N$ observables $A_{1}$, $A_{2}$, $\ldots$, $A_{N}$, we have the following standard deviation-based sum uncertainty relation,
\begin{equation}\label{cb3}
\sum_{i=1}^{N}\Delta A_{i}\geq\frac{1}{N-2}\left[\sum_{1\leq i<j\leq N}\Delta(A_{i}+A_{j})-\Delta\left(\sum_{i=1}^{N}A_{i}\right)\right].
\end{equation}

See Methods for the proof of Theorem 2.

If the lower bound (\ref{cb3}) is zero, then all $\Delta(A_{i}+A_{j})$ are equal to zero
(This can be seen next from the fact that our bound (\ref{cb3}) is tighter than $\Delta(\sum_{i=1}^{N}A_{i})$).
In this case, the measured state $|\psi\rangle$ is a common eigenstate of all the $N$ observables.
Hence standard deviation-based uncertainty inequality (\ref{cb3}) implies that the lower bound is nontrivial
whenever the $N$ observables are incompatible associated to the state. Therefore,
the standard deviation-based sum uncertainty relations also play the roles in characterizing the incompatibility of observables.

The lower bounds for sum uncertainty inequalities have been also provided in several arguments \cite{Huang,Rivas,Pati}.
In \cite{Pati}, the authors proved that for arbitrary $N$ observables $A_{1},A_{2},\ldots,A_{N}$, the sum of standard deviations of $N$ observables
is no less than the standard deviation of sum of the observables \cite{Pati},
\begin{equation}\label{tb2}
\sum_{i=1}^{N}\Delta A_{i}\geq\Delta\left(\sum_{i=1}^{N}A_{i}\right).
\end{equation}
Nevertheless, by using the following inequality,
$$
\begin{array}{rcl}
\displaystyle\sum_{1\leq i<j\leq N}\|a_{i}+a_{j}\|&\geq&\displaystyle\left\|\sum_{1\leq i<j\leq N}(a_{i}+a_{j})\right\|\\[5mm]
&=&\displaystyle(N-1)\left\|\sum_{i=1}^{N}a_{i}\right\|,
\end{array}
$$
one can show that our lower bound (\ref{cb3}) is tighter than (\ref{tb2}) in general.

\begin{figure}
\centering
\includegraphics[width=7cm]{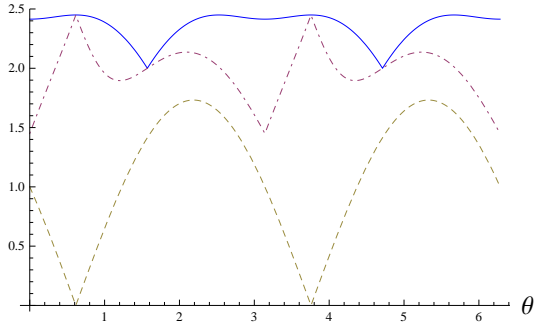}
\caption{The top solid line is the sum of the standard deviations $\Delta X+\Delta Y+\Delta Z$.
It can be reached by our lower bound (\ref{cb3}) (dash-dotted line) at $\theta=0.61548$, $\frac{\pi}{2}$, $0.61548+\pi$ and $\frac{3\pi}{2}$.
The dashed line stands for the bound (\ref{tb2}). The maximal value of the bound (\ref{tb2}) is $\sqrt{3}\approx1.732$,
while the bound (\ref{cb3}) can achieve its maximum 2.44949, which is equal to the actual sum uncertainties $\Delta X+\Delta Y+\Delta Z$
at $\theta=0.61548$ and $0.61548+\pi$.}\label{sd1}
\end{figure}

To compare the standard deviation-based sum uncertainty relation (\ref{cb3}) with the variance-based one (\ref{cb1}), let us
consider again the family of pure states given by the Bloch vector
$\overrightarrow{r}=(\frac{1}{\sqrt{2}}\cos\theta,\frac{1}{\sqrt{2}}\cos\theta,\sin\theta)$.
It is shown in FIG \ref{sd1} that the sum of standard deviations $\Delta X+\Delta Y+\Delta Z$ can attain the lower bound (\ref{cb3}),
while the variance-based sum uncertainties cannot reach the bound (\ref{cb1}), see FIG \ref{v1}.

\begin{figure}
\centering
\includegraphics[width=7cm]{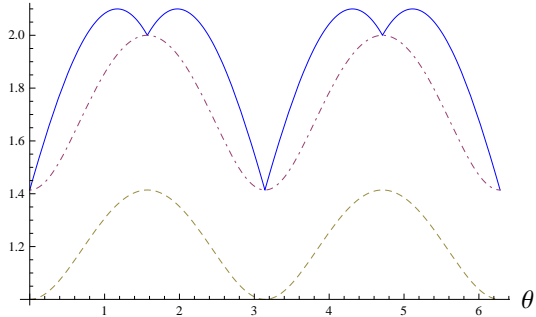}
\caption{Our lower bound (\ref{cb3}) (dash-dotted line) is tight to $\Delta J_{x}+\Delta J_{y}+\Delta J_{z}$ (solid line),
and they are equal when $\theta=0$, $\frac{\pi}{2}$, $\pi$ and $\frac{3\pi}{2}$.
The bound (\ref{cb3}) is always greater than the bound (\ref{tb2}) (dashed line) in this case.}\label{sd2}
\end{figure}

We have considered the uncertainty relations from measuring a qubit system by the spin-1/2 operators.
There are many physical systems of higher spin or angular momentum.
As another example, let us consider spin one systems.
Let $|\psi\rangle=\cos\frac{\theta}{2}|0\rangle+\sin\frac{\theta}{2}|2\rangle,~0\leq\theta<2\pi$
be a qutrit pure state. We choose three angular momentum operators ($\hbar=1$):
\begin{equation*}
J_{x}=\frac{1}{\sqrt{2}}\begin{pmatrix}
0 & 1 & 0\\1 & 0 & 1\\0 & 1 & 0
\end{pmatrix},~~~
J_{y}=\frac{1}{\sqrt{2}}\begin{pmatrix}
0 & -i & 0\\i & 0 & -i\\0 & i & 0
\end{pmatrix},~~~
\end{equation*}
\begin{equation*}
J_{z}=\begin{pmatrix}
1 & 0 & 0\\0 & 0 & 0\\0 & 0 & -1
\end{pmatrix}.
\end{equation*}
We have $(\Delta J_{x})^{2}=\frac{1}{2}(1+\sin\theta)$, $(\Delta J_{y})^{2}=\frac{1}{2}(1-\sin\theta)$, $(\Delta J_{z})^{2}=\sin^{2}\theta$,
$[\Delta(J_{x}+J_{y})]^{2}=1$, $[\Delta(J_{y}+J_{z})]^{2}=\frac{1}{2}(1-\sin\theta)+\sin^{2}\theta$,
$[\Delta(J_{x}+J_{z})]^{2}=\frac{1}{2}(1+\sin\theta)+\sin^{2}\theta$, $[\Delta(J_{x}+J_{y}+J_{z})]^{2}=1+\sin^{2}\theta$.
The sum of the standard deviations uncertainty relations are shown in FIG \ref{sd2}.
As the state $|\psi\rangle$ is not a common eigenstate of all the three angular momentum operators,
both inequalities (\ref{cb3}) and (\ref{tb2}) are not trivial. From Figs. \ref{sd1} and \ref{sd2}, it is also obvious that
our bound is tight.

\medskip
\noindent{\bf Conclusion}

We have provided two uncertainty relations for $N$ observables based on sum of variances and standard deviations, respectively.
Both uncertainty inequalities are useful in characterizing the incompatibility of arbitrary finite number of observables,
in the sense that the lower bounds are nontrivial as long as the measured state is not a common eigenstate of all the observables.
We have compared the variance-based with the standard deviation-based sum uncertainty relations by detailed examples of spin-1/2 systems.
A good lower bound must be a tighter one and has a clear physical implication.
Our results could also shed some light on applications of the uncertainty relation such as in entanglement detection \cite{Hofmann,G¨¹hne,Huang}.

\medskip
\noindent{\bf Methods}

{\sf Proof of Theorem 1}~To prove the inequality (\ref{cb1}), we need the following identity in a Hilbert space:
$$
\left\|\sum_{i=1}^{N}a_{i}\right\|^{2}+(N-2)\sum_{i=1}^{N}\|a_{i}\|^{2}=\sum_{1\leq i<j\leq N}\|a_{i}+a_{j}\|^{2},
$$
where $a_{i}$ are any vectors in the corresponding vector space, $\|\cdot\|$ stands for the norm of a vector defined by inner product.
Note that
\begin{equation}\nonumber
\begin{split}
\left\|\sum_{i=1}^{N}a_{i}\right\|&=\left\|\frac{1}{N-1}\sum_{1\leq i<j\leq N}(a_{i}+a_{j})\right\|\\
&\leq\frac{1}{N-1}\sum_{1\leq i<j\leq N}\|a_{i}+a_{j}\|,
\end{split}
\end{equation}
we have
\begin{equation}\label{ineqcb1}
\begin{split}
\sum_{i=1}^{N}\|a_{i}\|^{2}\geq&\frac{1}{N-2}\left[\sum_{1\leq i<j\leq N}\|a_{i}+a_{j}\|^{2}\right.\\
&\left.-\frac{1}{(N-1)^{2}}\left(\sum_{1\leq i<j\leq N}\|a_{i}+a_{j}\|\right)^{2}\right].
\end{split}
\end{equation}
Let $a_{i}=(A_{i}-\langle A_{i}\rangle)|\psi\rangle$, then $\|a_{i}\|=\Delta A_{i}$, and $\|a_{i}+a_{j}\|=\Delta( A_{i}+A_{j})$.
Substituting the above relations to the inequality (\ref{ineqcb1}), we obtain (\ref{cb1}) for any pure states $|\psi\rangle$.
For mixed states $\rho$, we only need to set $a_{i}=(A_{i}-\langle A_{i}\rangle_{\rho})S$, where $S$ is the square root of $\rho$, $\rho=S^{2}$.
This completes the proof.

\medskip
{\sf Proof of Theorem 2}~By using the generalized Hlawka's inequality \cite{Horn,Honda},
$$
\left\|\sum_{i=1}^{N}a_{i}\right\|+(N-2)\sum_{i=1}^{N}\|a_{i}\|\geq\sum_{1\leq i<j\leq N}\|a_{i}+a_{j}\|,
$$
and setting $a_{i}=(A_{i}-\langle A_{i}\rangle)|\psi\rangle$ for a pure state $|\psi\rangle$, or setting
$a_{i}=(A_{i}-\langle A_{i}\rangle_{\rho})\sqrt{\rho}$ for a mixed state $\rho$, we get (\ref{cb3}) directly.

\newpage
\bigskip
\noindent{\sf Acknowledgements}

\noindent The work is supported by the NSFC under number 11275131.

\bigskip
\noindent{\sf Author contributions}

\noindent  B.C. and S.-M.F. wrote the main manuscript text. Both of the authors reviewed the manuscript.

\bigskip
\noindent{\sf Additional Information}

\noindent Competing Financial Interests: The authors declare no competing financial interests.

\end{document}